\documentstyle [12pt,amsfonts] {article}
         \input epsf

\newcommand{\beq}{\begin{equation}}
\newcommand{\eeq}{\end{equation}}
\newcommand{\beqs}{\begin{eqnarray}}
\newcommand{\eeqs}{\end{eqnarray}}

\catcode`@=11
\@addtoreset{equation}{section}
\@addtoreset{equation}{subsection}
\def\theequation{\ifnum\value{section}=0 \arabic{equation}\ignorespaces
\else \ifnum\value{section}=-1 A.\arabic{equation}\ignorespaces
\else \ifnum\value{subsection}=0 \thesection.\arabic{equation}\ignorespaces
\else \thesection.\arabic{subsection}.\arabic{equation}\ignorespaces
                           \fi
                      \fi
                 \fi}
\catcode`@=12

\begin{document}

\baselineskip 6.0mm

\vspace{4mm}

\begin{center}

{\Large \bf Spanning Trees on Graphs \\
 and    Lattices in $d$ Dimensions}

\vspace{8mm}

\vspace{4mm}
{Robert Shrock}

{C. N. Yang Institute for Theoretical Physics, State University of New York,
Stony Brook, NY 11794 and Brookhaven National Laboratory, Upton, NY  11973}
 \vspace{4mm}

{and}

\vspace{4mm}

{F. Y. Wu}

{Department of Physics, Northeastern University, Boston, Massachusetts 02115}

\vspace{8mm}

{\bf Abstract}
\end{center}

The problem of enumerating spanning trees on graphs and lattices is considered.
We obtain bounds on the number of spanning trees $N_{ST}$ and establish
inequalities relating the numbers of spanning trees of different graphs or
lattices.  A general formulation is presented for the enumeration of spanning
trees on lattices in $d\geq 2$ dimensions, and is applied to the hypercubic,
body-centered cubic, face-centered cubic, and specific planar lattices
including the kagom\'e, diced, 4-8-8 (bathroom-tile), Union Jack, and 3-12-12
lattices.  This leads to closed-form expressions for $N_{ST}$ for these
lattices of finite sizes.  We prove a theorem concerning the classes of graphs
and lattices ${\cal L}$ with the property that $N_{ST} \sim \exp(nz_{\cal L})$
as the number of vertices $n \to \infty$, where $z_{\cal L}$ is a finite
nonzero constant.  This includes the bulk limit of lattices in any spatial
dimension, and also sections of lattices whose lengths in some dimensions go to
infinity while others are finite.  We evaluate $z_{\cal L}$ exactly for the
lattices we considered, and discuss the dependence of $z_{\cal L}$ on $d$ and
the lattice coordination number.  We also establish a relation connecting
$z_{\cal L}$ to the free energy of the critical Ising model for planar lattices
${\cal L}$.
 
\vspace{16mm}

\pagestyle{empty}
\newpage

\pagestyle{plain}
\pagenumbering{arabic}
\renewcommand{\thefootnote}{\arabic{footnote}}
\setcounter{footnote}{0}

\section{Introduction}

The enumeration of spanning trees on a graph or lattice is a problem of
long-standing interest in mathematics \cite{fh} - \cite{cvet} and physics
\cite{kirk} - \cite{wu77}.  Let $G=(V,E)$ denote a connected graph (without
loops) with vertex (site) and edge (bond) sets $V$ and $E$.  Let $n=v(G)=|V|$
be the number of vertices and $e(G)=|E|$ the number of edges.  A spanning
subgraph $G^\prime$ is a subgraph of $G$ with $v(G^\prime) = |V|$, and a tree
is a connected subgraph with no circuits.  A spanning tree is a spanning
subgraph of $G$ that is a tree (thus $e(G') = n-1$).  The degree of a vertex is
the number of edges attached to it (often denoted coordination number or
valence). A $\kappa$-regular graph is a graph with the property that each of
its vertices has the same degree $\kappa$.  For these and further related
definitions see, e.g., \cite{fh} - \cite{welsh}.

Denote the number of distinct spanning trees of a graph $G$ by $N_{ST}(G)$.
This number can be enumerated in terms of standard graph-theoretic quantities.
Two methods for doing this will be used here: (i) via the Laplacian matrix
\cite{fh,bbook,cvet}, and (ii) as a special value of the Tutte polynomial
\cite{tutte1} - \cite{tutte3}.  For the first, we recall the definition that
two vertices are adjacent if they are connected by an edge. The adjacency
matrix ${\bf A}$ of $G$ is an $n \times n$ matrix whose elements are 
\beqs
A_{ij}&=& 1, \hskip 1cm {\rm if\>\>sites\>\>}i\>\>{\rm and}\>\>j\>\>{\rm are
\>\>adjacent} \nonumber \\ &=& 0, \hskip 1cm {\rm otherwise}, 
\label{adj}
\eeqs 
and the
degree matrix ${\bf \Delta}$ of $G$ is an $n\times n$ diagonal matrix with
elements 
\beq 
\Delta_{ij} = \kappa_i \delta_{ij}\ , 
\label{Delta} 
\eeq 
where
$\kappa_i$ is the degree of site $i$, and $\delta_{ij}$ the Kronecker delta
function. Define the Laplacian matrix 
\beq 
{\bf Q} = {\bf \Delta} - {\bf A} \ .
\label{matrixQ}
\eeq
Here and throughout this paper, we use boldface to denote matrices of size
$n\times n$.

Since the sum of the elements in each row (or column) of ${\bf Q}$ vanishes,
 one of the eigenvalues of ${\bf Q}$ is zero.  Denote the remaining $n-1$
 eigenvalues by $\lambda_1,...,\lambda_{n-1}$.  Then two basic theorems in
 graph theory state that \cite{fh,bbook,cvet} 
\beqs 
N_{ST}(G)&=& {\rm Any\>\>cofactor\>\>of\>\>} {\bf Q} \label{eigen}\\ &=&
 \frac{1}{n}\prod_{i=1}^{n-1}\lambda_i \ .
\label{muprod}
\eeqs
An elementary proof of the equivalence of (\ref{eigen}) and (\ref{muprod}) can
be found in \cite{wust}.  The Laplacian matrix ${\bf Q}$ is also known in the
literature as the Kirchhoff matrix, or simply the tree matrix, which arose in
the analysis of electric circuits \cite{kirk}.  The enumeration of $N_{ST}$
has very recently  been considered in \cite{wust} for finite hypercubic 
lattices and the square net embedded on nonorientable surfaces.

A second way to calculate $N_{ST}(G)$ is as the special value
\beq
N_{ST}(G)=T(G,1,1)
\label{t11}
\eeq
of the Tutte polynomial of the graph $G$ \cite{tutte1} - \cite{tutte3}, 
\beq
T(G,x,y) = \sum_{G^\prime \subseteq G} 
(x-1)^{k(G^\prime)-k(G)}(y-1)^{c(G^\prime)},
\label{t}
\eeq
where
$k(G)$ is the number of connected components and $c(G)$ the number of 
independent circuits in $G$, and 
 the summation is over all spanning subgraphs $G'$ of $G$.
Here, we have $k(G)=1$ for connected graphs $G$, and
 it is clear that (\ref{t}) leads to  (\ref{t11}), since in the limit 
of $x,y \to 1$ the only contributing terms 
 in (\ref{t}) are those of 
$k(G^\prime)=k(G)=1$ and $c(G^\prime)=0$, namely, the spanning 
trees. 

In physics one often deals with lattices.
A lattice is regular if all sites are equivalent \cite{gs}.
  For a wide class of graphs, including lattices which may or may not be
regular  and strips of lattices, 
the number of spanning trees $N_{ST}$ has the asymptotic exponential growth
\beq
N_{ST}(G) \sim \exp(nz_{\{G\}}) \quad {\rm as} \quad n \to \infty\ .
\label{nexp}
\eeq
Thus $z_{\{G\}}$  provides a natural measure of the rate of growth, and is
evaluated via 
\beq
z_{\{G\}} = \lim _{n\to \infty} {1\over n} \ln N_{ST}(G) \ . 
\label{zg}
\eeq
where $\{G\}$ denotes the formal  $n\to\infty$ limit of a graph of type $G$.
For  lattices ${\cal L}$ this is known as the bulk, or the thermodynamic,
limit, and we denote $z_{\{G\}}$ by $z_{\cal L}$.  Closed-form 
expressions for $z_{\cal L}$ have been obtained for the square, 
honeycomb, and triangular lattices \cite{temprel,wu77}.  Exact results have
also been obtained for strips of regular lattices of finite widths and  infinite length \cite{gh} - \cite{ta}.  

\vspace{8mm}

In the present work we consider spanning trees on general graphs $G$ as well as in the limit 
of $x,y \to 1$
 lattices ${\cal L}$ in $d\geq 2$ dimensions.  Specifically, we shall

(i) derive an exact relation between $z_{\cal L}$ and $z_{{\cal L}^*}$ 
for planar lattices ${\cal L}$ and ${\cal L}^*$ 
which are mutually dual.
 
(ii) obtain  bounds on $N_{ST}(G)$ and $z_{\{G\}}$,

(iii) establish the exponential growth
 $N_{ST}(G) \sim \exp(nz_{\{G\}} )$ for a wide class of  graphs,

(iv) present a general formulation for the enumeration of 
  $N_{ST}({\cal L})$ and $z_{\cal L}$, 

(v) enumerate $N_{ST}({\cal L})$ and evaluate $z_{\cal L}$ exactly for a number
of lattices in $d\geq 2$,

(vi) analyze the dependence of $z_{\cal L}$ on the spatial dimensionality and
    the coordination number of the lattice ${\cal L}$, including deriving an
    asymptotic expansion for $d$-dimensional hypercubic lattices, and

(vii) establish a relation connecting $z_{\cal L}$ to the free energy of the
    critical Ising model for planar lattices ${\cal L}$.

\vspace{8mm}

Before proceeding, it is useful to review here the close connection of spanning
trees with the Potts model in statistical mechanics The partition function of a
$q$-state Potts model at temperature $T=1/\beta$ on $G$ is \cite{potts,wurev} 
\beq
Z(G,q,v)=\sum_{\{\sigma_i\}}e^{-\beta {\cal H}}
\label{zpotts}
\eeq
where 
$-\beta {\cal H} = K\sum_{\langle ij \rangle} \delta_{\sigma_i \sigma_j}$
 with 
$\langle ij \rangle$ ranging over pairs of adjacent 
vertices in $G$.  The summation is taken over $\sigma_i=1,2,...,q$ and 
$i=1,2,...,n$.
The partition function (\ref{zpotts}) can be written as \cite{whit,kf}
\beq 
Z(G,q,v) = \sum_{G^\prime \subseteq G} q^{k(G^\prime)}v^{e(G^\prime)},
\label{zqv}
\eeq
where the summation is over all spanning subgraphs of $G$ and 
\beq
v=e^K-1 \ . 
\label{vdef}
\eeq
The expression (\ref{zqv}) shows that $Z(G,q,v)$ is a polynomial in $q$ and $v$
and enables one to generalize $q$ from
positive integers to real and complex values.  The Potts model partition 
function on a graph $G$ is related to the Tutte polynomial according to
\beq
Z(G,q,v) = (x-1)^{k(G)}(y-1)^{n}T(G,x,y)
 \label{ztutte}
\eeq
with
\beq
x=1+q/v \ , \quad y=v+1\
\label{xydef}
\eeq
so
\beq
q=(1-x)(1-y) \ . 
\label{qxy}
\eeq
Thus, $N_{ST}(G)$ can also be calculated in terms of the Potts partition
function $Z(G,q,v)$.  The values $x=1$ and $y=1$ in the evaluation (\ref{t11})
correspond to the limits $q \to 0$, and $v \to 0$ with 
$\lim_{\{q,v\} \to 0}(q/v) = 0$.  

For a planar graph $G$ and its  dual graph $G^*$,
the Tutte polynomial possesses the duality relation \cite{bbook,tutte2,tutte3}
\beq
T(G,x,y)=T(G^*,y,x).
\label{tuttedual}
\eeq
An immediate consequence of (\ref{t11}) and (\ref{tuttedual}) 
is that the number of spanning trees is the same for a planar graph $G$ and 
its dual $G^*$ (see, for example, \cite{bbook}), namely, 
\beq
N_{ST}(G)=N_{ST}(G^*) \ . 
\label{ndual}
\eeq

Let $n^*$ be the number of sites of $G^*$, given by the Euler relation
\beq
n^* = |E|-n+1\ .
\label{euler}
\eeq
For planar lattices ${\cal L}$ and its dual  ${\cal L}^*$,
it is convenient to introduce the ratio in the bulk limit
\beq
\nu_{\cal L} = \lim _{n\to\infty} ({{n^*}/ n}) 
\label{nu}
\eeq
  satisfying
\beq
\nu_{\cal L}\nu_{{\cal L}^*}=1 \  .
\label{nudual}
\eeq
Using (\ref{nu}) and (\ref{nudual}), we obtain the relation
%derive the following relation between
%$z$ on a regular planar lattice ${\cal L}$ and the dual lattice ${\cal L}^*$:
 \beq
z_{{\cal L}^*} =  z_{\cal L}/\nu_{\cal L} \ 
\label{ratio}
\eeq
relating $z_{{\cal L}}$ and $ z_{{\cal L}^*}$.
 As examples, it is readily verified that we have
\beq
 \nu_{hc}=1/2, \hskip 1cm \nu_{kag} = 1, \hskip 1cm \nu_{4-8-8}=1/2\ ,
\eeq
where the subscripts denote
 the honeycomb, kagom\'e,  and 4-8-8 lattices, respectively
(for a detailed discussion on classifications of planar lattices see, for 
example, \cite{gs,wn}).
Applying (\ref{ratio}) to the respective duals, namely, the
triangular, diced, and Union Jack  lattices,   we obtain the relations
   \beq
z_{tri} = 2\ z_{hc}\ , \hskip 1cm 
z_{diced} = z_{kag}\ , \hskip 1cm 
 z_{UJ} = 2\ z_{4-8-8}. 
\label{trihc}
\eeq

For a regular planar lattice involving a tiling of the plane with only one type
of polygon, the dual lattice is also regular.
   This  includes the square which is self-dual, and
honeycomb and triangular lattices which are mutually dual.  In
contrast, for a regular planar lattice involving a tiling of the plane with
more than one type of polygon, the  dual lattice is not regular
since it does not have a uniform coordination number.  For example, the
kagom\'e and 4-8-8 lattices, which are regular, involve   tilings with more than one type
of polygon, and their duals, the diced and Union Jack lattices, are not
regular.  For nonregular lattices it is convenient to introduce an
effective coordination number $\kappa_{eff}$ defined as the average number of
edges per vertex, 
\beq
\kappa_{eff} = \lim_{n \to \infty}({2|E|}/{n}) \ .   
\label{eff}
\eeq
Combining (\ref{euler})  and (\ref{eff}), we derive the relation 
\beq
\kappa_{eff} = 2(1+\nu_{\cal L}) \ . 
\label{effk}
\eeq
As examples, using (\ref{nudual}) this 
yields $\kappa_{eff}= 4$ and 6, respectively, for the diced and the
Union Jack lattices.  For $\kappa$-regular graphs we have  
$\kappa_{eff}=\kappa$.

\section{General Bounds} 

The determination of upper bounds on $N_{ST}(G)$ is a problem of considerable
interest in graph theory.  A general upper bound is \cite{grim}
\beq
N_{ST}(G) \le \frac{1}{n}\biggl ( \frac{2|E|}{n-1}\biggr )^{n-1},
\label{kboundg}
\eeq
For a $\kappa$-regular graph, this implies the upper bound
\beq
N_{ST}(G) \le \frac{1}{n}\biggl ( \frac{n\kappa}{n-1} \biggr )^{n-1} 
\label{kboundgregular}
\eeq
and hence
\beq
z_{\cal L} \leq \ln \kappa \ .
\label{b1}
\eeq 
More generally, we shall also deal with lattices that are not
regular but for which one can define an effective coordination number,
$\kappa_{eff}$ as in   (\ref{eff}). For these, from (\ref{kboundg}), one has
\beq 
z_{\cal L} \leq \ln \kappa_{eff} \ .
\label{b1nonreg}
\eeq

A stronger upper bound for $\kappa$-regular graphs with $\kappa \ge 3$ is due to 
Mckay, Chung, and Yau \cite{mckay,cy}, who established rigorously that
 \beq 
N_{ST}(G) \le \Biggl ( \frac{2\ln n}{n \kappa \ln \kappa} 
\Bigg) (C_\kappa) ^n, 
\label{nmckay}
\eeq
where
\beq
C_\kappa = \frac{(\kappa-1)^{\kappa-1}}{[\kappa(\kappa-2)]^{\kappa/2-1}} \ .
\label{ck}
\eeq
This leads to the upper bound 
\beqs
z_{\cal L} &\leq& \ln C_\kappa  
\label{zmckay} \\
 &=& \ln \kappa - \biggl [ \frac{1}{2\kappa} + \frac{1}{2\kappa^2} +
 \frac{7}{12\kappa^3} 
+ \frac{3}{4\kappa^4} + \frac{31}{30\kappa^5} +\frac{3}{2\kappa^6} + 
O\Bigl ( \frac{1}{\kappa^7} \Bigr ) \biggr ]   \ , 
\label{zupperklarge}
 \eeqs
where we have carried out  a large-$\kappa$ expansion.

On the other hand, lower bounds on $N_{ST}(G)$ are more difficult to obtain.
We have established an inequality which we state as a theorem:
 
\begin{flushleft}

Theorem 2.1.

Let $G$ be a connected graph and let $i$ and $j$ be two nonadjacent
vertices of $G$ with degrees $\kappa_i$ and $\kappa_j$, respectively. 
Let $H$ be a graph obtained from $G$ by adding an edge $e_{ij}$ connecting 
$i$ and $j$. Then we have
\beq
N_{ST}(H) > \Big( {{\kappa+1}\over \kappa}\Big) N_{ST}(G),
 \eeq
where $\kappa = {\rm min} \{\kappa_i, \kappa_j\}$.

\vspace{6mm}

Proof: Since sites $i$ and $j$ are connected in $G$, the adding of the edge
$e_{ij}$ to any spanning tree $T(G)$ on $G$ forms a closed circuit on $H$.  The
closed circuit contains in addition to $e_{ij}$ another edge $\ell$ incident at
site $j$. The deletion of $\ell$ then breaks the circuit, resulting in a
spanning tree configuration $T(H)$ on $H$.  However, the spanning tree $T(H)$
so constructed is not necessarily unique; the same $T(H)$ may result
from $m$ different $T(G)$.  Since each $T(G)$ is also a spanning tree of $H$,
it follows that for the $m$ spanning trees $T(G)$ there exist $m+1$ distinct
spanning trees $T(H)$.  A moment's reflection shows that we have always $m\leq
\kappa $, with $m=\kappa$ arising when there is a single edge incident at site
$j$ in $T(G)$ and a single edge $e_{ij}$ at $j$ in $T(H)$.  Since we have
$(\kappa_j+1)/\kappa_j \geq (\kappa_i+1)/\kappa_i$ for $\kappa_i\geq \kappa_j$,
the proposition follows as a consequence. $\Box$

\vspace{6mm}

Corollary 2.1.

Let $G$ be a $\kappa$-regular graph, and let $G'$ be a graph
derived from $G$ by adding $M$ edges one at a time such that each added edge
terminates in at least one vertex whose degree is $\kappa$.  Then 
\beq
N_{ST}({\cal L}') > \Big( {{\kappa+1}\over \kappa}\Big)^{M} N_{ST}({\cal L}).
\label{boundL}
\eeq
 
\vspace{6mm}

Remark: Corollary 2.1 is proved by applying Theorem 2.1 $M$ times.
Furthermore, for  lattices $z_{{\cal L}}$ and $z_{{\cal L}'}$, and
$M=\alpha n$ where $\alpha$ is a constant, (\ref{boundL}) implies the bound 
\beq 
z_{{\cal L}'} > z_{{\cal L}}+\alpha \ln \Big( {{\kappa+1}\over \kappa}\Big) 
\ . 
\label{zlow} 
\eeq
\end{flushleft} 
For example, by adding edges one at a time one can convert the
honeycomb lattice first to the square and then to the triangular lattice.  
Corollary 2.1 then implies the inequalities 
\beq
z_{sq} > z_{hc} + {1\over 2}\ln \Bigg({4\over 3}\Bigg) 
\label{zsqbgen}
\eeq
and
\beq
z_{tri} > z_{sq} + \ln\Bigg({5\over 4}\Bigg) \ . 
\label{ztribgen}
\eeq
Combining these bounds with the relation
$z_{tri} = 2 z_{hc}$ from (\ref{trihc}), we obtain the lower bounds 
\beqs
 z_{sq}
&>& \ln \Big( {5\over 3} \Big) = 0.510\ 825\ 6...
\label{zsqb} \\
 z_{hc}& >& \ln \Big( {5\over {2 \sqrt 3}} \Big) = 0.366\ 984\ 5...
\label{zhcb} \\
 z_{tri} &> &2 \ln \Big( {5\over {2 \sqrt 3}} \Big) = 0.733\ 969\ 1...
\label{ztrib}
\eeqs
Note that these bounds have been deduced without actually carrying out 
explicit calculations.  For comparison, the exact values are
\cite{temprel,wu77} (see also Sec. 4 below)
\beq
z_{sq} = \frac{4}{\pi}\biggl [1- \frac{1}{3^2} + \frac{1}{5^2} - 
\frac{1}{7^2} + \frac{1}{9^2} - ... \biggr ] = 1.166\ 243\ 6...
\label{zsq}
\eeq
and \cite{wu77} 
\beqs
z_{hc}&=& \frac{3\sqrt{3}}{2\pi}\biggl [ 1 - \frac{1}{5^2} + \frac{1}{7^2} 
- \frac{1}{11^2} + \frac{1}{13^2} - ... \biggr ] = 0.807\ 664\ 9...
\label{zhc} \\
 z_{tri} &=& 2z_{hc} =  1.615\ 329\ 7...
\label{ztri}
\eeqs

To measure the effectiveness of these lower bounds,
let $R_{\{G\}}$ denote the ratio of the exact $z_{\{G\}}$ to the  
bounds (\ref{zsqb}) - (\ref{ztrib}).  We have
\beq
R_{sq} \simeq 2.28 , \hskip 1cm 
 R_{hc} = R_{tri} \simeq 2.20 \ , 
\label{rhclow}
\eeq
indicating that the  bounds   are very generous. 

It is also of interest to work out the implications of the lower bound 
(\ref{zlow}) for strips of regular lattices. For example, for 
$2 \times \infty$ square \cite{gh,a} and triangular \cite{ta} strips
  one has the results 
\cite{gh,a,ta} 
\beqs
z^{free}_{sq(2 \times \infty)}&=&\frac{1}{2}\ln(3+\sqrt{2}  \ ) 
= 0.658\ 478\ 9... \nonumber \\
z^{free}_{tri(2 \times \infty)} &=& \frac{1}{2}\ln \biggl ( \frac{7+3\sqrt{5}}{2} 
\biggr ) = 0.962\ 423\ 6...\ ,
\label{zsqly2}
\eeqs
where the superscript denotes free boundary conditions in the
transverse direction.  This is to be compared with  (\ref{zsqn6})
and (\ref{ztrin6}) below for periodic boundary conditions in the transverse direction
for which we have 
$z_{sq(2 \times \infty)} = 0.881\ 373\ 5... $ and 
$z_{tri(2 \times \infty)} =  1.386\ 294\ 3...$
It is easily checked that these numbers obey (\ref{ztribgen}).
 Furthermore, the inequality (\ref{ztribgen}) now implies
\beqs
&&z^{free}_{tri(2 \times \infty)} > z^{free}_{sq(2 \times \infty)}+
 \frac{1}{2}\ln \Bigl ( \frac{4}{3} \Bigr )
  = 0.802\ 319\ 9... \nonumber \\
&&z_{tri(2 \times \infty)} > z_{sq(2 \times \infty)}+
 \frac{1}{2}\ln \Bigl ( \frac{4}{3} \Bigr )
  = 1.025\ 214\ 5...
\label{ztrisq}
\eeqs
so that the ratios of the exact values to the lower bounds (\ref{ztrisq}) are
\beq
R^{free}_{tri(2 \times \infty)} \simeq 1.20, \hskip 1cm
R_{tri(2 \times \infty)} \simeq 1.35 \ . 
\label{rsqly2low}
\eeq
% These ratios are considerably closer to unity, i.e. the respective lower bounds
% are considerably better than is the case with (\ref{rhclow}). 

Another example is the $d$-dimensional hypercubic lattice ${\cal
L}_d$ with $\kappa=2d$ for which Corollary 2.1 
implies\footnote{\footnotesize{Here we add a technical remark.
As the inequality (\ref{yd}) is deduced by
constructing ${\cal L}_{d+1}$ via adding edges to connect $n^{1/d}$ copies of $
{{\cal L}_{d}}$,  a technical problem arises if the copies are disjoint to
begin with for which $N_{ST}=0$ by definition.  The difficulty is resolved if
one starts instead from copies of $ {{\cal L}_{d}}$ which are connected by a
single edge.}}
  \beq 
z_{{\cal L}_{d+1}}>z_{{\cal
L}_d} +\ln \Big[ 1+(2d)^{-1}\Big]. 
\label{yd} 
\eeq 
Applying this lower bound to $z_{sq}$ using $z_{line}=0$, we
obtain $z_{sq} > \ln(3/2)$, which is not as strong as the bound
(\ref{zsqb}).  For the $d=3$ simple cubic lattice, we have
\beq
z_{sc} > z_{sq} + \ln \Bigl ( \frac{5}{4} \Bigr ) = 1.389\ 387\ 1...
\label{zsc}
\eeq
so that 
\beq
R_{sc} \simeq 1.20\ .
\label{scell}
\eeq
Similar lower bounds can be deduced for 
hypercubic lattices with $d \ge 4$.

\section{Families of Graphs with Exponential Growth for $N_{ST}(G)$}

In this section we  prove a result concerning the class of families of
graphs for which $N_{ST}(G)$ has the exponential asymptotic behavior 
(\ref{nexp}).  

A family of graphs is recursive if
it can be built up by  sequential additions of a given subgraph.  As an
example, consider a strip of width $N_2$   
and length $N_1=m$
of some  lattice such as a square lattice; this can be built up by
starting with a column of $N_2-1$ squares and sequentially adding columns to
elongate the strip in the $N_1$ direction. A higher-dimensional example is
a rectangular tube of a  lattice such as a simple cubic lattice with
transverse size $N_2 \times N_3$ and length $N_1=m$; this can be built up by
starting with a single $N_2 \times N_3$ section and sequentially adding $m$
transverse sections and connecting them in an obvious manner to elongate the tube.  
% The boundary conditions in the various directions may be free or
% periodic. The total number of vertices is then given by
% \beq
% n = t\ m + t_0
% \label{ntau}
% \eeq
% where $t$ and $t_0$ are integers  whose values
% depend on the type of lattice, the size of the transverse
% section, and the boundary conditions. 
  We will need the following result from \cite{a}:

\begin{flushleft} 

\vspace{4mm}

Lemma 3.1.

 Let $G_m$ be a recursive graph of length $m$ subunits as described
in the above.
   Then the Tutte polynomial has the form 
\beq
T(G_m,x,y) = \sum_{j=1}^{N_a} c_{G,j}(x,y) (a_{G,j}(x,y))^m\ ,
\label{tgform}
\eeq
where  the explicit forms of $a_{G,j}(x,y)$ and $c_{G,j}(x,y)$ 
  depend
on the type of graphs $G_m$.  

\vspace{4mm}

The proof of Lemma 3.1 can be found in \cite{a}.  (See (2.18) and (8.15) of
\cite{a}, where $a_{G,j}$ was denoted by $\lambda_{G,j}$.)
 % $\Box$
Note that the class of recursive graphs is more general than the class of
regular lattices. 

\end{flushleft}
\vspace{4mm}
 
  As discussed in
\cite{a},  as $m \to \infty$ for a given $(x,y)$, the term $a_{G,j}$ with the
maximal magnitude will dominate the right-hand side of (\ref{tgform}), provided
that the corresponding coefficient $c_{G,j}(x,y)$ does not vanish. We denote this
term $a_{G,j,max}$.  Using the relation between the Potts partition function
and the Tutte polynomial (\ref{ztutte}), together with the definition of the
(reduced) Potts model free energy 
\beq
f(\{G\},q,v) = \lim_{n \to \infty} n^{-1} \ln Z(G,q,v)\ ,
\label{f}
\eeq
one observes that a nonanalyticity in $f$ occurs when, as one changes $(x,y)$
or equivalently $(q,v)$, there is a crossover of  the dominant term $a_{G,j}$.
These changes determine the regions of analyticity (phases) of the free
energy. 
We next proceed to our theorem.
 
\begin{flushleft}
\vspace{4mm}

Theorem 3.1.

 Let $G_m$ denote a recursive graph of a  lattice 
of length $m$ in one spatial dimension with fixed
$(d-1)$-dimensional ``transverse'' section of the 
size $N_2\times N_3\times \cdot\cdot
N_d$.   For $N_2,\cdot\cdot\cdot,N_d \ge 2$, the number of spanning trees 
$N_{ST}(G_m)$ grows exponentially with $n$ as in (\ref{nexp}),
 thereby defining a nonzero finite constant $z_{\{ G\}}$. 

\vspace{8mm}

Proof.  \ We shall carry out the proof for the case $d=2$; the generalization
to $d \ge 3$ is straightforward.  The strategy of the proof is to use the
structural result (\ref{tgform}) in  Lemma 3.1 above.  It is convenient to
use some results from the Potts-Tutte correspondence (\ref{ztutte}).  From
(\ref{vdef}) and (\ref{xydef}) it follows that $y=1$ corresponds, in terms of
 the correspondence (\ref{ztutte}), to the infinite-temperature point $K=v=0$. 
 From the basic property that a spin model such as the Potts model
is analytic at $K=0$, i.e., has a Taylor series expansion in $K$ (or
$v$) with a finite radius of convergence, it follows that,
 in the
neighborhood of $y=1$ for a given $x$,  a single term 
\beq
a_{G,max}=a_{G,R_1}
\label{agmax}
\eeq
will dominate,
where $R_1$ denotes the region in the $(x,y)$ space corresponding to the 
paramagnetic phase
for a given $q$ in the $(q,v)$ space.  It was shown in \cite{a} that the
corresponding coefficient in $T(G,x,y)$ is nonzero. Further, one knows that
$a_{G,max}(1,1) > 1$ unless $G$ is the tree or circuit graph or obvious
modifications thereof, a fact which
 can be easily  proved by assuming the contrary and
deducing that $N_{ST}$ violates a lower bound on the number of spanning trees
for strip graphs with width $N_2 \ge 2$. From   (\ref{t11}) and
(\ref{zg}), it follows that $N_{ST}(G_m)$ has the exponential asymptotic growth
as in (\ref{nexp}), so that there is a finite nonzero 
constant $z_{\{ G\}}$.  The method of the proof evidently provides a constructive
way to calculate this quantity in terms of (\ref{tgform}), or 
\beq 
z_{\{ G\}} = t^{-1} \ln [a_{G,j,max}(1,1)]\ ,
\label{zstripa}
\eeq
where $t$ is the number of vertices in a transverse section.
 It is straightforward to generalize these results to the case of a
$d$-dimensional tube graph with fixed $(d-1)$-dimensional transverse section. 
This completes the proof.  \ $\Box$

\vspace{4mm}

\end{flushleft}

To place this result in perspective, we give some examples of families of
graphs for which the asymptotic behavior (\ref{nexp}) does not hold.  For the
tree graph $T_n$ and circuit graph $C_n$, the Tutte polynomials are
$T(T_n,x,y)=x^{n-1}$ and $T(C_n,x,y)=y+\sum_{s=1}^{n-1}x^s$ so that
$N_{ST}(T_n)=1$, independent of $n$, and $N_{ST}(C_n)=n$.   In both cases
$z_{\{T\}}=z_{\{C\}}=0$.  It is for this reason that we restricted the width of
the strips to  $N_2,N_3,\cdots \ge 2$ in Theorem 3.1. 
An example of a family for which
$N_{ST}$ grows more rapidly than an exponential is the complete graph $K_n$,
a graph with the property  that each vertex is adjacent to every other vertex.
In this case, one has $N_{ST}(K_n)=n^{n-2}$ \cite{bbook}.  Note that $K_n$ is
not a recursive graph.
We next proceed to the case of the  bulk limit of a 
lattice.

\begin{flushleft}
 \vspace{4mm}

Theorem 3.2.

Let ${\cal L}$ denote the bulk limit of a regular $d$-dimensional 
lattice of size $N_1 \times ... \times N_d$, with $N_j \to
\infty$, $j=1,...,d$ such that the ratios $\lim_{n \to \infty} N_i/N_j$ remain
 nonzero and finite.  Then the number of spanning trees on this lattice 
grows exponentially with $n$ as in (\ref{nexp}),
 thereby defining a nonzero finite constant $z_{\cal L}$.

\vspace{4mm}

Proof.  A sketch of the proof is given here.  The idea is to use Theorem 3.1 
 and observe that the property of exponential growth of $N_{ST}$ 
as $N_1 \to \infty$ for fixed $N_2$ is independent of $N_2\ge 2$.
 Furthermore,
as discussed in the proof of Theorem 3.1, given the relation 
(\ref{t11}), it follows that $N_{ST}$ and $z$ are, in statistical mechanics
terminology, determined by $Z(G,q,v)$ and the free energy $f$ at the disorder 
point $v=K=0$. Hence, we can take the limit $N_2 \to \infty$, and
since the exponential growth of $N_{ST}$ holds uniformly in $N_2$, it also 
holds in this limit.  This establishes the result for  $d=2$.  It is
straightforward to extend the proof to  $d \ge 3$ by using Theorem 3.1
starting with a tube of the $d$-dimensional lattice with
fixed $(d-1)$-dimensional transverse cross section and one longitudinal
direction that goes to infinity, and using again the fact that the exponential
growth of $N_{ST}$ holds uniformly as one increases the $(d-1)$ transverse
dimensions of the tube.  $\Box$

\vspace{5mm}

\end{flushleft}

We remark that Theorem 3.2 can also be established directly using the explicit
expression (\ref{general}) of $N_{ST}({\cal L})$  obtained in the next section.
Another remark is that it is important that $z_{\cal }$ is
  a disorder quantity.  Because of this, one
gets the same asymptotic behavior for $z$ on a sequence of infinite-length 
tube graphs of progressively larger and larger transverse cross sections as one
does by taking the limit of all  $N_j\to\infty$  with
the ratios $N_i/N_j$ fixed.    In contrast, this
would not be the case in dealing with a quantity connected
with a  divergent correlation length.  For example, the free energy of the
ferromagnetic Potts model on infinite-length, finite-width strips has a 
zero-temperature critical point at $v \to \infty$ for any finite width, but has a quite different
analytic structure if one takes both $N_1$ and $N_2$ to infinity with $N_2/N_1$
fixed, namely, a non-analyticity (phase transition)
at a finite temperature $v > 1$.

\section{Formulation for General Lattices}
The formulation of enumerating spanning trees for general lattices is given in
this section.  Consider a lattice ${\cal L}$ of $n$ sites in $d$ spatial
dimensions.  We shall use (\ref{muprod}) to evaluate $N_{ST}$ and for
simplicity assume periodic boundary conditions.  Formulations for other
boundary conditions can be similarly worked out (see, for example,
\cite{wust}).

To write down the
Laplacian matrix ${\bf Q}$ in a form suitable for computing its eigenvalues, we
make use of the fact that any  lattice in $d$ dimensions is decomposable
into a hypercubic array of $N_1\times N_2\times \cdots \times N_d$ unit cells,
each containing $\nu$ sites so that we have $n=\nu N_1N_2\cdots 
N_d$.\footnote{\footnotesize{If $N_j=2$ for some $j$, 
  then the two sites in the $j$-th direction
are connected by double edges.
But this does not affect any of   the   ensuing discussions.}}
 Specify
the cells by the coordinate ${\bf n}=\{n_1, n_2, \cdots, n_d\}$, where $n_i=0,
1, 2, \cdots, N_i-1$, and number the sites in a cell $1, 2,\cdots,\nu$.  Let
$a({\bf n}, {\bf n}')$ be the $\nu\times\nu$ cell vertex adjacency matrix 
describing the connectivity between the vertices of the unit cells 
${\bf n}$ and ${\bf n}'$. Namely,
\beqs
a_{ij}({\bf n}, {\bf n}')&=& 1, \hskip 1cm 
{\rm if\>\>site\>\>}i{\rm \>\> in\>\> cell\>\> {\bf n} \>\>and \>\>
site\>\>}j{\rm \>\>in\>\>cell\>\> {\bf n}'\>\>are\>\> adjacent} \nonumber \\
&=& 0, \hskip 1cm {\rm otherwise}.
\eeqs
Under the assumption of periodic boundary conditions, we have the 
translational symmetry
\beq
 a({\bf n}, {\bf n}')= a({\bf n}-{\bf n}'),
\eeq
and we can therefore write $a({\bf n})=a(n_1, n_2, \cdots, n_d)$.

\begin{figure}[hbp]
        \centering
        \leavevmode
        \epsfxsize=5.5in
        \begin{center}
        \leavevmode
        \epsffile{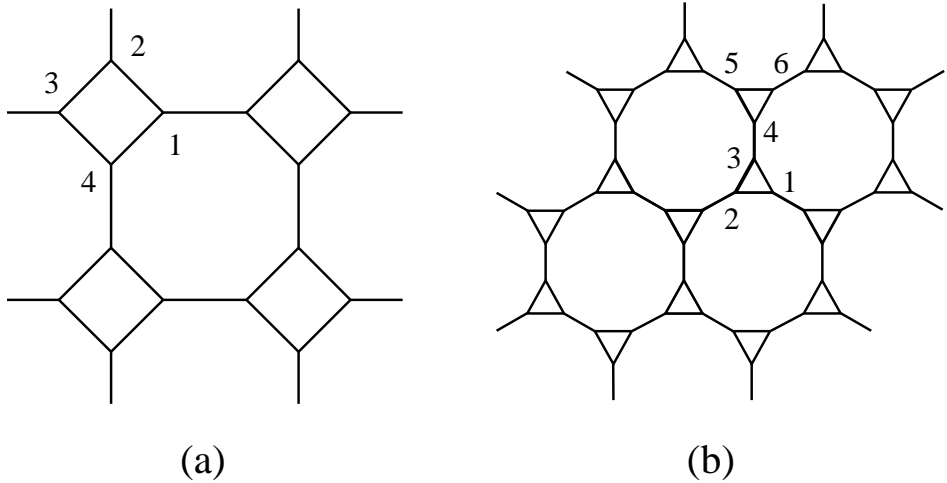}
        \end{center}
        \vspace{-4.5in}
        \caption{\footnotesize{(a) The 4-8-8 (bathroom tile) lattice. (b)  The
        3-12-12 lattice. Sites within a unit cell are labeled as shown.}}
         \label{fig488}
       \end{figure}

The general formulation is best illustrated by considering an example.  Here we
consider the 4-8-8 (bathroom-tile) lattice shown in Fig. \ref{fig488}(a).  This 
is a regular lattice which has the
coordination number 3, and the unit cells are the obliquely 
oriented squares, with $\nu=4$.  The explicit forms for the $a(n_1,n_2)$ 
matrices depend on one's convention for labeling the vertices
within a unit cell; we choose the labeling shown in Fig. \ref{fig488}(a).
Then one has
 
\beqs
a(0,0)&=& \left(\begin{array}{cccc}
0&1&0&1\\
1&0&1&0\\
0&1&0&1\\
1&0&1&0
 \end{array}
\right),
\hskip 1cm 
a(1,0)= a^T(-1,0)=\left(\begin{array}{cccc}
0&0&1&0\\
0&0&0&0\\
0&0&0&0\\
0&0&0&0
 \end{array}
\right), \nonumber \\
a(0,1)&=& a^T(0,-1)=\left(\begin{array}{cccc}
0&0&0&0\\
0&0&0&1\\
0&0&0&0\\
0&0&0&0
 \end{array}
\right),
 \eeqs
where $a^T$ is the transpose of $a$.
Then the Laplacian matrix ${\bf Q}$ assumes the form
\beqs
{\bf Q} &=&  \Big[3I_4 -a(0,0)\Big] \otimes I_{N_1} \otimes I_{N_2}
  - a(1,0)\otimes R_{N_1}\otimes I_{N_2} + a(-1,0
)\otimes R^T_{N_1}  \otimes I_{N_2}\nonumber \\
  && \hskip 2cm - a(0,1)\otimes I_{N_1} \otimes R_{N_2}
- a(0,-1)\otimes I_{N_1} \otimes R^T_{N_2},
\eeqs
where $I_N$ is the $N \times N$ identity matrix, and 
$R_N$  the $N \times N$  matrix
\beq
R_N=\left(\begin{array}{ccccc}
0&1&0&\cdots&0\\
0&0&1&\cdots&0\\
\vdots&\vdots&\vdots&\ddots&\vdots\\
0&0&0&\cdots&1\\
1&0&0&\cdots&0
\end{array}
\right).
\eeq 

To determine the  eigenvalues of  ${\bf Q}$, we make use of the fact  that
$R_N$ is 
diagonalized by  the  similarity transformation $S_NR_NS_N^{-1}$ 
generated by the  matrix $S_N$ with elements 
\beq
\left(S_N\right)_{nm} =\left(S_N^{-1}\right)^*_{mn} 
=N^{-1/2}e^{i2\pi mn/N}, \hskip 1cm m,n=0,1,\cdots, N-1,
\eeq
where $^*$ denotes the complex conjugate, yielding  the 
 eigenvalues 
\beq
\lambda_n= e^{i2\pi n/N} , \hskip 1cm n=0,1,\cdots,N-1.  \label{eigenper}
\eeq
 It follows that  the similarity transformation generated by
\beq
I_{\nu}  \otimes S_{N_1} \otimes  S_{N_2} 
\eeq
diagonalizes ${\bf Q}$ in the $N_1$ and $N_2$ subspaces.  
 Then, using the fact that
a determinant is equal to the product of its eigenvalues, we obtain
from (\ref{muprod})  the expression
\beq
N_{ST}({\cal L}_{4-8-8}) = \Bigg({{\Lambda}
\over {4N_1N_2}}\Bigg) \prod_{ k_1=0}^{N_1-1}
\prod_{ k_2=0}^{N_2-1} \det \Bigl | M \Bigg( {{2\pi k_1}\over {N_1}} , 
{{2\pi k_2}\over {N_2}} \Bigg)\Bigr | , 
\hskip 0.5cm (k_1,k_2)\not=(0,0), \label{48}
\eeq
where $\Lambda =\lambda_1\lambda_2\lambda_3=64$, and
$ \lambda_{1}=\lambda_{2}=\lambda_{3} =4$ are  the nonzero eigenvalues
of $M(0,0)$.  Here, $M$ is a $4\times 4$ matrix
and the notation ${\rm det} |M|$   denotes the determinant of $M$.  Explicitly,
we have
 \beqs
M(\theta_1,\theta_2) &=& 3I_4 -a(0,0)-a(1,0)e^{i\theta_1}
 -a(-1,0)e^{-i\theta_1} 
% \nonumber \\
% && \hskip 2cm
 -a(0,1)e^{i\theta_2}-a(0,-1)e^{-i\theta_2} \nonumber \cr\cr
&=&\left(\begin{array}{cccc}
3&-1&-e^{i\theta_1}&-1\\
-1&3&-1&-e^{i\theta_2}\\
-e^{-i\theta_1}&-1&3&-1\\
-1&-e^{-i\theta_2}&-1&3
\end{array} \right)
\label{m488}
\eeqs
and hence 
\beq
 \det |M(\theta_1,\theta_2)| = 4\ \Big[
7-3(\cos\theta_1+\cos\theta_2) - \cos\theta_1\cos\theta_2 \Big] \ .
\label{detm488}
\eeq
Note that a change in labeling conventions would either interchange 
or negate $\theta_1$ 
and/or $\theta_2$.  These have no effect on the final expression  since 
${\rm det}|M|$ is invariant under these changes.
   From (\ref{detm488}) we have 
\beqs
z({\cal L}_{4-8-8})
&=&  {1\over 2} \ln 2 + {1\over {4} } \int_{-\pi}^{\pi}
{{d\theta_1}\over {2\pi}}\int_{-\pi}^{\pi} {{d\theta_2}\over {2\pi}} 
\ln \Big[7-3(\cos\theta_1+\cos\theta_2) - \cos\theta_1\cos\theta_2 \Big] 
\nonumber \cr\cr
&=& {1\over 4} \ln 2 +{1\over {4\pi} } \int_{0}^{\pi}d\theta
 \ln\Big[7-3\cos\theta + 4\sin(\theta/2) \sqrt{5-\cos\theta} \Big] \nonumber \\
  &=& 0.786\ 684(1). \label{48z}
\eeqs
The last two lines are obtained after carrying out one integration
followed by a numerical integration of the remaining integral.
 
\medskip 

The consideration of  general ${\cal L}$  now proceeds  in a similar fashion.
 In place of (\ref{48}), one obtains
\beq
N_{ST}({\cal L}) = \Bigg({{\Lambda}
\over {\nu N_1\cdot\cdot N_d}} \Bigg)\prod_{ k_1=0}^{N_1-1}\cdot\cdot
\prod_{ k_d=0}^{N_d-1} D\Bigg( {{2\pi k_1}\over {N_1}}, 
\cdots,{{2\pi k_d}\over {N_d}}\Bigg)  \hskip 0.5cm
 ({\bf k} \not={\bf 0}).  
\label{general}
\eeq
where ${\bf k}=(k_1,k_2,\cdots,k_d)$, ${\bf 0} = (0,0,\cdots,0)$,
\beqs
D(\theta_1,\cdots,\theta_d)& =& {\rm det}\ \big| M(\theta_1,\cdots,\theta_d) \big|\ ,
\label{determinant} \\
 \Lambda &=&1, \hskip 2.4cm \nu=1 \nonumber \\
&=& \lambda_1 \cdot\cdot \lambda_{\nu-1}, \hskip 1cm \nu>1\ ,
\eeqs
and $\lambda_i$'s are the $\nu-1$ nonzero eigenvalues of the matrix 
$M(0,\cdot\cdot,0)$.  
Here, $M$ is a $\nu\times \nu$ matrix defined by 
\beq
M(\theta_1,  \cdots,\theta_d)
=  \Delta_\nu - \sum_{\bf n} a ({\bf n}) e^{i {\bf n}\cdot { \bf \Theta}}\ ,
\label{mdef}
\eeq
where $\Delta_\nu$ is the degree matrix (\ref{Delta}) for a unit cell,
and $\Theta=(\theta_1,\cdot\cdot, \theta_d)$. 
Note that the determinant of $M$ is always real since the 
 matrix $M$ is hermitian.
 This leads to the result 
\beq
z_{\cal L} \equiv z({\cal L})  =   {1\over {\nu} } \int_{-\pi}^{\pi}{{d\theta_1}\over {2\pi}}
\cdots \int_{-\pi}^{\pi} {{d\theta_d}\over {2\pi}} 
\ln  D(\theta_1,\cdots,\theta_d) \ . 
\label{main}
\eeq
 Eqs. (\ref{general}) and  (\ref{main}) are our main results for general 
regular lattices, and (\ref{general}) is suitable for enumerating
$N_{ST}$ for lattices of finite sizes.

It is also of interest to consider the case in which the size of the lattice 
is finite in $\ell < d$ dimensions and goes to infinity in the remaining
$d-\ell$ dimensions.  For example, if one sets $\ell=1$, then as $N_1$
increases from 1 to infinity, the resultant sequence of values of $z$ can be
regarded as a sort of ``interpolation'' between the infinite 
$(d-1)$-dimensional and infinite
$d$-dimensional lattices.  Without the loss of generality we let 
$L_1, \cdots L_\ell$ be finite.  Then from (\ref{general}) we have 
\beqs 
&& z \Bigl ( {\cal L}(N_1\times \cdot\cdot \times N_\ell \times \infty \times
\cdot\cdot \times \infty) \Bigr ) = \frac{1}{\nu N_1\cdot\cdot N_\ell} \times 
\nonumber \\
 && \quad \times \sum_{k_{1}=0}^{N_1-1} \cdot\cdot
\sum_{k_\ell=0}^{N_\ell-1} \int_{-\pi}^\pi  {{d\theta_{\ell+1}}\over {2\pi}}
 \cdot\cdot  \int_{-\pi}^\pi  {{d\theta_d}\over {2\pi}}
 D \Bigg(
 {{2\pi k_{1}}\over {N_{1}}}, \cdot\cdot,{{2\pi k_\ell}\over {N_\ell}},
 \theta_{\ell+1}, \cdot\cdot \theta_d  \Bigg )  
 \ . 
 \label{zstrip}
 \eeqs
This shows clearly the result of Theorem 3.1, that $N_{ST}$  grows 
exponentially like (\ref{nexp}). For  $d=2, \ell =1$, i.e., the case of 
$N\times \infty$ strips of infinite length and finite width $N_1 \equiv N$,
the integration can be carried out.  In particular,  for
 strips of square and triangular lattices, we obtain the expressions
\beqs
z_{sq(N\times \infty)}& =& \frac{1}{N}\sum_{k=0}^{N-1} \ln \biggl [ 
2-\cos\omega_k + \biggl ( \Bigl (2-\cos\omega_k \Bigr )^2 -1 
\biggr )^{1/2} \ \biggr ]
\label{zsqstrip} \\
 z_{tri( N\times \infty )} &=& \frac{1}{N}\sum_{k=0}^{N-1} \ln \biggl [
3-\cos\omega_k + \biggl ( (1-\cos\omega_k)(7-\cos\omega_k) 
\biggr )^{1/2} \ \biggr ]\ ,
\label{ztristrip}
\eeqs
where  $\omega_k =2\pi k/N$.
 The summations  
(\ref{zsqstrip}) and (\ref{ztristrip}) can be explicitly carried out
for a given $N$. We give the results  in Sec. 6.6 below.

\section{Lattices in $d \geq 3$ Dimensions} 

The formulation of the preceding section is now specialized to specific 
lattices in $d\geq 3$ dimensions.
\subsection{$d$-Dimensional Hypercubic Lattices}
Consider first the $d$-dimensional hypercubic lattice ${\cal L}_d$ consisting
of $n=N_1N_2\cdots N_d$ sites.  Here, ${\cal L}_2$ and ${\cal L}_3$ are the
square and simple cubic lattices. We have $\nu=1$, $\Delta_1=\kappa = 2d$, and the 
adjacency matrices 
\beqs
a({\bf n}) &=&1,\hskip 1cm 
 {\bf n} = (\pm 1, 0, \cdots, 0), (0,\pm 1,  \cdots, 0),
\cdots, (0,0, \cdots, \pm 1) \nonumber \\
 &=& 0, \hskip 1cm {\rm otherwise}.
\eeqs
Using the general expressions (\ref{general}) and  (\ref{main}), one obtains immediately
\beq
N_{ST}({\cal L}_d) ={{2^{n-1}}\over n} \prod_{ k_1=0}^{N_1-1} \cdots
 \prod_{ k_d=0}^{N_d-1}\bigg [ d-\bigg(\cos \Bigl ( {{2 k_1\pi}\over N_1} 
\Bigr ) +\cdots + \cos \Bigl ( {{2 k_d\pi}\over N_d} \Big) \biggr ) 
\biggr ] ,  \>\> ({\bf k} \not= {\bf 0}) 
\label{gen}
\eeq
 From   (\ref{gen}) we derive 
\beq
z({\cal L}_d) = \ln (2 d)+ \int_{-\pi}^{\pi}{{d\theta_1}\over {2\pi}}
\cdots \int_{-\pi}^{\pi}{{d\theta_d}\over {2\pi}}
\ln \Big[ 1-{1\over d}(\cos\theta_1 +\cdots +\cos\theta_d)\Big] \ .
\label{latveclamd}
\eeq
The expression (\ref{gen}) is the same as that reported in \cite{wust}.
 
As in the case of the 4-8-8 lattice,  one of the integrations can be carried 
out analytically and the remaining $(d-1)$-fold integral done numerically.
For $d=2$, the quantity $z({\cal L}_2)$ can be exactly evaluated 
\cite{temprel,wu77}, leading to the value given by
(\ref{zsq}) which we include in  (\ref{list}) below.
We have carried out the  
numerical integrations for $d=3,4$ and present the results also
 in (\ref{list}). For higher values of $d$ the
evaluation of the integral becomes less accurate; however one
can expand the logarithm in (\ref{latveclamd}) and carry out the 
integrations term by term.  This leads to the large-$d$ expression
\beq
z({{\cal L}_d}) = \ln(2d) -\biggl [ \frac{1}{4d} + 
\frac{3}{16d^2} + \frac{7}{32d^3} + \frac{45}{128d^4} + 
\frac{269}{384d^5} + \frac{805}{512d^6} + O\Bigl (\frac{1}{d^7} \Bigr ) 
\biggr ] \ .
\label{zlarge_d}
\eeq
Note that the expansion (\ref{zlarge_d}) agrees with the corresponding
large-$\kappa$ expansion (\ref{zupperklarge}) of the Mckay-Fan-Yau upper bound 
 to the order of $d^{-1}$, after taking $\kappa=2d$.
However, $1/d$ expansions are generally expected to be
asymptotic in view of the slow growth of the coefficients  (see, for example,
\cite{gf}).  If one truncates the   $1/d$ series to a given 
order and lets $d \to \infty$, the approximation to 
 $z({{\cal L}_d})$ becomes progressively more accurate.
But for a fixed $d$, one does not necessarily obtain a more accurate 
approximation by including more terms in the calculation.  
In practice, however, we  found that, for $d=4,5,$ and 6, the 
series evaluated to $O(d^{-4})$ and $O(d^{-5})$ gives essentially 
 the same values, which we listed in (\ref{list}),
with an accuracy of $10^{-4}$. Combining our results, we have
  \beqs
z_{sq}= z({\cal L}_2)&=&  1.166\ 243\ 6\cdots \nonumber \\
z({\cal L}_3)&=& 1.674\ 148\ 1(1) \hskip 1.5cm {\rm (numerical\>\>evaluation)} \nonumber \\
z({{\cal L}_4})&=& 2.000\ 0(5)\hskip 2.3cm {\rm (numerical\>\>evaluation)} \nonumber \\
z({\cal L}_5)&=& 2.243 \hskip 3.2cm {\rm (series\>\>expansion)}\nonumber \\
z({\cal L}_6)&=&2.437\hskip 3.2cm {\rm (series\>\>expansion)}\nonumber \\
z({\cal L}_d) &\to& \ln (2d), \hskip 1cm d\to \infty.\label{list}
\eeqs
Our numerical result suggests that $z({{\cal L}_4})$ may be exactly equal to 2.
It is readily verified that values of $z({\cal L}_d)$ in (\ref{list}) are
consistent with the inequality (\ref{yd}).
 
\subsection{$d$-Dimensional Body-Centered Cubic Lattice}
For the usual three-dimensional body-centered cubic (bcc) lattice, a unit cell
 contains $\nu=2$ vertices located at $(0,0,0)$ and $({1\over 2},{1\over
 2},{1\over 2})$ numbered 1 and 2, respectively.  Then one has $\Delta_2=8I_2$
 and the adjacency matrices
\beqs
a(0,0,0)&=&\pmatrix{0&1\cr 1&0\cr} \cr\cr
a(1,0,0) &=&a(0,1,0) =a(0,0,1) =a(1,1,0) \cr\cr
 &=&a(1,0,1) =a(0,1,1) =a(1,1,1) =\pmatrix{0&1\cr 0&0\cr} \cr\cr
a(-1,0,0) &=&a(0,-1,0) =a(0,0,-1) =a(-1,-1,0) \cr\cr
 &=&a(-1,0,-1) =a(0,-1,-1) =a(-1,-1,-1) =\pmatrix{0&0\cr 1&0\cr}, 
\eeqs
leading to the matrix
\beq
M(\theta_1,\theta_2) = \pmatrix{8& -v_1v_2v_3 \cr -(v_1v_2v_3)^* & 8\cr}
\eeq
where $v_j=1+e^{i\theta_j}$, $j=1,2,3$.  
The evaluation of the determinant yields
\beq
D(\theta_1,\theta_2) = 64\Big[ 1-\cos^2(\theta_1/2)
 \cos^2(\theta_2/2)\cos^2(\theta_3/2) \Big] .
\eeq
This leads to 
\beqs
N_{ST}({\cal L}_{bcc}) &=&\Bigg({{\Lambda}\over {n}}\Bigg) 
\prod_{ k_1=0}^{N_1-1}
\prod_{ k_2=0}^{N_2-1}
 \prod_{ k_3=0}^{N_3-1}\Bigg[ 64-64\cos^2 \Big({{ k_1\pi}\over N_1} \Big)
  \cos^2 \Big({{ k_2\pi}\over N_2}\Big) \cos^2 \Big({{ k_3\pi}\over N_3}\Big)
\Bigg], \label{bccfinite} \nonumber \\
&& \hskip 5.5cm 
(k_1,k_2,k_3)\not=(0,0,0) 
\eeqs
where $n=2N_1N_2N_3,$ and $ \Lambda=16$ is the nonzero eigenvalue of $M(0,0)$.
The expression (\ref{bccfinite}) enumerates  $N_{ST}$ for finite
bcc lattices. Using 
%the  integration formula
%\beq
%\frac{1}{2}\int_{-\pi}^{\pi} \cdots \int_{-\pi}^{\pi}
%\frac{d^d\theta_j}{(2\pi)^d} \ \ln (1-\prod_{j=1}^d \cos^2(\theta_j/2) ) = 
%\int_{-\pi}^{\pi} \cdots \int_{-\pi}^{\pi}
%\frac{d^d\theta_j}{(2\pi)^d} \ \ln (1-\prod_{j=1}^d \cos \theta_j )
%\label{intform}
%\eeq
%together with 
(\ref{zg}), we obtain 
\beqs
 z({\cal L}_{bcc})& = & 3\ln 2 + 
\frac{1}{2}\int_{-\pi}^{\pi}{{d\theta_1}\over {2\pi}}
\int_{-\pi}^{\pi}{{d\theta_2}\over {2\pi}}
 \int_{-\pi}^{\pi}{{d\theta_3}\over {2\pi}}
\ln \Bigg[ 1-\cos^2\Bigg({{\theta_1}\over 2}\Bigg)
\cos^2\Bigg({{\theta_2}\over 2}\Bigg)
\cos^2\Bigg({{\theta_3}\over 2}\Bigg)
 \Bigg]  \nonumber \\
& =& 3\ln 2 + \int_{-\pi}^{\pi}{{d\theta_1}\over {2\pi}}
\int_{-\pi}^{\pi}{{d\theta_2}\over {2\pi}}
 \int_{-\pi}^{\pi}{{d\theta_3}\over {2\pi}}
\ln \Big( 1-\cos\theta_1 \cos\theta_2 \cos\theta_3\Big) \ . 
\label{bcc3}
\eeqs
These results can be generalized to the $d$-dimensional body-centered cubic
lattice, which we shall denote bcc(d).  This lattice has coordination number
$\kappa = 2^d$. For finite lattices we obtain
 \beq
N_{ST}({\cal L}_{bcc(d)}) =
\Bigg({{\Lambda}\over {2N_1\cdot\cdot N_d}}\Bigg) \prod_{ k_1=0}^{N_1-1}
\cdot\cdot
\prod_{ k_d=0}^{N_d-1}
 \Bigg[ 2^{2d}-2^{2d}\prod_{j=1}^d \cos^2 \Big({{ k_j\pi}\over N_j}
 \Big)
\Bigg],\>\>({\bf k}\not={\bf 0})
 \eeq
where $\Lambda =2^{d+1}$. Taking the bulk limit, we obtain 
\beqs
z({\cal L}_{bcc(d)})&=&d\ \ln 2 + \int_{-\pi}^{\pi}{{d\theta_1}\over {2\pi}}
\cdots \int_{-\pi}^{\pi}{{d\theta_d}\over {2\pi}}
\ln \Big( 1-(\cos\theta_1 )\cdots (\cos\theta_d)\Big) \nonumber \\
&=& d\ \ln2 -\frac{1}{2}\sum_{\ell=1}^\infty \frac{1}{\ell}\Biggl (
\frac{(2\ell)!}{2^{2\ell}(\ell !)^2} \Biggr )^d \ , \label{bccd}
\eeqs
where we have expanded the logarithm and carried out the integrations term by
term.  

For $d=2$, the body-centered cubic lattice $bcc(2)$ is just the square
lattice, and it is readily seen that the two integrals in (\ref{latveclamd})
and (\ref{bccd}) are equal for $d=2$.  Interestingly, this also establishes the
equality of the two series in (\ref{zsq}) and (\ref{bccd}) at $d=2$.  We have
further evaluated $z({\cal L}_{bcc(d)})$ for $d\geq 3$ using both expressions
in (\ref{bccd}), and found that the series converges slowly, with good
agreement between the two reached only after evaluating the series to $100-200$
terms.  For $d=3$ and 4 the results are
\beqs 
z({\cal L}_{bcc})&=&1.990\ 2(1), \hskip 1.5cm d=3 \nonumber \\ 
z({\cal L}_{bcc(4)})&=& 2.732\ 3(1), \hskip 1.5cm d=4.  
\eeqs

\subsection{Face-Centered Cubic Lattice}

For the face-centered cubic (fcc) 
lattice, a unit cell contains $\nu=4$ vertices located at
$(0,0,0)$, $(0,{1\over 2},{1\over 2}),$ $ ({1\over 2},0,{1\over 2}), ({1\over
2},{1\over 2},0)$ numbered $1,2,3,4$, respectively.  One has $\Delta_4 =
(12)I_4$ and the adjacency matrices,
\beqs
&&a(0,0,0)= \left(\begin{array}{cccc}
0&-1&-1&-1\\
-1&0&-1&-1\\
-1&-1&0&-1\\
-1&-1&-1&0
\end{array}
\right), \hskip 1cm 
a(1,0,0) =a^T(-1,0,0)=\left(\begin{array}{cccc}
0&0&0&0\\
0&0&0&0\\
1&1&0&0\\
1&1&0&0
\end{array}
\right) \nonumber \\
&&a(0,1,0) =a^T(0,-1,0)=\left(\begin{array}{cccc}
0&0&0&0\\
1&0&1&0\\
0&0&0&0\\
1&0&1&0
\end{array}
\right), \>\>
a(0,0,1) =a^T(0,0,-1)=\left(\begin{array}{cccc}
0&0&0&0\\
1&0&0&1\\
1&0&0&1\\
0&0&0&0
\end{array}
\right) \nonumber\\
&&a(1,1,0) =a^T(-1,-1,0)=\left(\begin{array}{cccc}
0&0&0&0\\
0&0&0&0\\
0&0&0&0\\
1&0&0&0
\end{array}
\right),\>\>
a(1,0,1) =a^T(-1,0,-1)=\left(\begin{array}{cccc}
0&0&0&0\\
0&0&0&0\\
1&0&0&0\\
0&0&0&0
\end{array}
\right) \nonumber\\
&&a(0,1,1) =a^T(0,-1,-1)=\left(\begin{array}{cccc}
0&0&0&0\\
1&0&0&0\\
0&0&0&0\\
0&0&0&0
\end{array}
\right), \>\>
a(1,-1,0) =a^T(-1,1,0)=\left(\begin{array}{cccc}
0&0&0&0\\
0&0&0&0\\
0&1&0&0\\
0&0&0&0
\end{array}
\right) \nonumber\\
&&a(1,0,-1) =a^T(-1,0,1)=\left(\begin{array}{cccc}
0&0&0&0\\
0&0&0&0\\
0&0&0&0\\
0&1&0&0
\end{array}
\right),\>\>
a(0,1,-1) =a^T(0,-1,1)=\left(\begin{array}{cccc}
0&0&0&0\\
0&0&0&0\\
0&0&0&0\\
0&0&1&0
\end{array}
\right) ,  \cr\cr
 &&\label{fcca}
\eeqs
leading to the matrix
\beq
M(\theta_1,\theta_2,\theta_3) =
\left(\begin{array}{cccc}
12&-(v_2v_3)^*&-(v_1v_3)^*&-(v_1v_2)^*\\
-v_2v_3&12&-v_1^*v_2&-v_1^*v_3\\
-v_1v_3&-v_1v_2^*&12&-v_2^*v_3 \\
-v_1v_2&-v_1v_3^*&-v_2v_3^*&12
\end{array}
\right) , 
\label{fccM}
\eeq
where $v_j=1+e^{i\theta_j}, j=1,2,3$. The evaluation of the determinant yields
\beqs
D(\theta_1,\theta_2,\theta_3) &=&
12^4\  F(\theta_1,\theta_2,\theta_3) \nonumber \\
 F(\theta_1,\theta_2,\theta_3)&=&  1-{2\over 9} (c_1+c_2+c_3) 
-{8\over {27}} c_1c_2c_3 -{2\over {81}} c_1c_2c_3(c_1+c_2+c_3) \nonumber \\
&& \hskip 2cm  
 +{1\over {81}}
(c_1^2+c_2^2+c_3^2) , \label{fccD}
 \eeqs
where $c_i=\cos ^2(\theta_i/2)$.
This gives
\beq
N_{ST}({\cal L}_{fcc}) =\Bigg({{\Lambda}\over n}\Bigg)\prod_{ k_1=0}^{N_1-1}
\prod_{ k_2=0}^{N_2-1}
 \prod_{ k_3=0}^{N_3-1} D\Big( {{ 2k_1\pi}\over N_1},  {{2 k_2\pi}\over N_2},
{{ 2k_3\pi}\over N_3} \Big),
 \>\>(k_1,k_2,k_3)\not=(0,0,0) 
\eeq
where $\Lambda=16^3,n=4N_1N_2N_3$, so that 
\beqs
z_{fcc} & = & \ln (12) +\frac{1}{4} \int_{-\pi}^{\pi}{{d\theta_1}\over{2\pi}}
 \int_{-\pi}^{\pi}{{d\theta_2}\over{2\pi}} \int_{-\pi}^{\pi}{{d\theta_3}\over{2\pi}}
 \ln  F(  \theta_1,\theta_2,\theta_3).
\label{zfcc}
\eeqs
  The numerical evaluation of (\ref{zfcc})
   yields the value
\beq
z_{fcc}=2.354\ 4(4).
\label{zffvalue}
\eeq

\section{Planar Lattices} 
In this section the formulation of section 4 is applied to some other planar
lattices (the square and 4-8-8 lattices have been considered in preceding
sections).

\subsection{Triangular Lattice}
The triangular lattice can be regarded as an 
$N_1\times N_2$ square net of sites with one additional diagonal edge added, in
the same way, to every square of the net. In this picture we have $\nu=1$,
$a(\pm 1,0)=a(0,\pm 1) = a(1,1) = a(-1,-1) =1$, and 
\beq
M(\theta_1,\theta_2) =6 - (e^{i\theta_1} + e^{-i\theta_1} + e^{i\theta_2} + 
e^{-i\theta_2} + e^{i(\theta_1+\theta_2)} + e^{-i(\theta_1+\theta_2)} ) \ . 
\eeq
It follows that
\beqs
z_{tri}&=&\int_{-\pi}^{\pi}
\frac{d\theta_1 }{2\pi}\int_{-\pi}^{\pi}
\frac{d\theta_2 }{2\pi} \ln \Bigl [ 6- 2\Bigl ( \cos\theta_1 + \cos\theta_2 + 
\cos(\theta_1 +\theta_2) \Bigr ) \Bigr ] \cr\cr
&=& \frac{3\sqrt{3}}{\pi}(1 - 5^{-2} + 7^{-2} - 11^{-2} + 
13^{-2} - ...) \cr\cr
&=& 1.615\ 329\ 736...
\label{tri}
\eeqs
The result (\ref{tri}) was reported previously in \cite{wu77} where it was
derived  via the connection to the Potts 
partition function and a mapping to a solvable vertex model.  For a finite
triangular lattice of $N_1\times N_2$ cells, the number of spanning trees
$N_{ST}$ is given by (\ref{general}) with $d=2,\Lambda= \nu=1$, an expression
we shall not reproduce here.

\subsection{Honeycomb Lattice}

It is instructive to derive $z_{hc}$ using the general formulation.  The
honeycomb lattice is a square net of unit cells of $\nu=2$ sites.  Consider the
honeycomb lattice in the form of a ``brick wall" and regard the two sites
connected by a vertical edges as forming a unit cell.  Then one has
\beq
a(0,0) = \pmatrix{0&1\cr 1&0\cr}, \>\>
a(1,0) =a(-1,0)= a^T(0,1)=a^T(0,-1)= \pmatrix{0&1\cr 0&0\cr},
\eeq
and
\beqs
M(\theta_1,\theta_2) &=& \pmatrix{3 & -(1+e^{i\theta_1}+e^{i\theta_2})\cr
               -(1+e^{-i\theta_1}+e^{-i\theta_2}) & 3 \cr}   \cr\cr
D(\theta_1,\theta_2)  &=&  6- 2\Bigl ( \cos\theta_1 +
\cos\theta_2 + \cos(\theta_1 +\theta_2) \Bigr ) \ . 
\eeqs
This leads to the relation $z_{hc} = z_{tri}/2$ given in (\ref{trihc}). 
For finite honeycomb lattices the number of spanning trees is the same as 
that of its dual, the triangular lattice.
  
\subsection{Kagom\'e and Diced Lattices}

The kagom\'e lattice has the structure of a square net of unit cells
each  containing $\nu=3$ sites forming a triangle. Therefore one has
 \beqs
a(0,0) &=&  \pmatrix{0&1&1 \cr 1&0&1\cr 1 & 1& 0\cr}, \hskip 2.5cm
a(1,0)= a^T(-1,0) = \pmatrix{0&0&0 \cr 0&0&0\cr 1 & 0& 0\cr} \cr\cr 
a(0,1)&= & a^T(0,-1) = \pmatrix{0&1&0 \cr 0&0&0\cr 0 & 0& 0\cr},\>\>
a(1,1)= a^T(-1,-1) = \pmatrix{0&0&0 \cr 0&0&0\cr 0 & 1& 0\cr}, 
\eeqs
and
\beq
M(\theta_1, \theta_2) = 
\pmatrix { 4 & -(1+e^{i\theta_2}) & -(1+e^{-i\theta_1}) \cr
  -(1+e^{-i\theta_2}) & 4  & -(1+e^{-i(\theta_1+\theta_2)}) \cr
  -(1+e^{i\theta_1})  & -(1+e^{i(\theta_1+\theta_2)}) & 4 \cr},
\eeq
with
\beq
D(\theta_1, \theta_2) = 12\Big[3-\Big(\cos\theta_1+
 \cos\theta_2 +\cos(\theta_1+\theta_2)\Big)\Big] \ .
\eeq
This yields the result 
\beq
z_{kag} =  (z_{tri}+\ln 6 )/3. \label{kagz}
\eeq
For a finite kagom\'e  lattice of  $N_1\times N_2$ cells, the number of
spanning trees $N_{ST}$ is given by (\ref{general}) with $d=2,\Lambda=6^2, 
\nu=3$.
 
For the diced lattice, which is the dual of the kagom\'e lattice, from 
(\ref{ndual}) and (\ref{trihc}), we obtain
\beq
N_{ST}({\cal L}_{diced})=N_{ST}({\cal L}_{kag}), \hskip 1cm 
z({\cal L}_{diced})=z({\cal L}_{kag}).
\eeq

\subsection{$3-12-12$ Lattice}
The 3-12-12 lattice is the lattice shown in Fig. \ref{fig488}(b)  which has 
the structure of  a square net with unit cells
each containing $\nu=6$ sites.  
   Label the six sites of a unit cell as shown,  one has
\beq
M(\theta_1,\theta_2) = \pmatrix{3&-1&-1&0&-e^{i\theta_1} & 0\cr
     -1&3&-1&0&0& -e^{-i\theta_2}\cr
         -1&-1&3&-1&0&0 \cr
        0&0&-1&3&-1&-1\cr
         -e^{-i\theta_1}&0&0&-1&3&-1 \cr
          0&-e^{i\theta_2}&0&-1&-1&3 \cr}, 
\eeq
with
\beq
D(\theta_1,\theta_2)  = 30\Big[ 3- \Bigl ( \cos\theta_1 + \cos\theta_2 +
\cos(\theta_1+\theta_2)\Bigr ) \Big] \ . 
\eeq
Hence, 
\beq
z({\cal L}_{3-12-12}) = {1\over 6}\Big[z_{tri}+\ln (15)\Big]\ .  
\eeq
For a finite 3-12-12 lattice of  $N_1\times N_2$ cells, the number of
spanning trees $N_{ST}$ is given by (\ref{general}) with $\Lambda=2\cdot 3^2
\cdot 5^2=450$ and $\nu=6$.

\subsection{Union Jack Lattice}
 
The Union Jack lattice  is the dual of the 4-8-8 lattice.  We  use 
(\ref{ndual}) and (\ref{trihc}) to obtain
\beq
N_{ST}({\cal L}_{UJ})=N_{ST}({\cal L}_{4-8-8}), \hskip 1cm
z({\cal L}_{UJ})=2\ z({\cal L}_{4-8-8}),\label{uj}
\eeq
where expressions of the right-hand sides in (\ref{uj}) are given
in, respectively,  (\ref{48}) and (\ref{48z}).

\subsection {$N\times \infty$ Lattice Strips}

In this section we give results on the explicit evaluation of $z$ for
$N\times \infty$ strips of square and triangular lattices for $2\leq N\leq 6$
by using (\ref{zsqstrip}) and (\ref{ztristrip}).  The results are
  \beqs
z_{sq(2 \times \infty)}&=&\frac{1}{2}\ln(3+2\sqrt{2} \ ) = 0.881\ 373\ 5...
\nonumber \\
 z_{sq(3 \times \infty)}&=&\frac{2}{3}\ln \Bigl ( \frac{5+\sqrt{21} \ }{2} \Bigr ) 
= 1.044\ 532\ 8...
\nonumber \\
 z_{sq(4 \times \infty)}&=& \frac{1}{4}\biggl [ \ln(3+2\sqrt{2} \ ) + 
2\ln(2+\sqrt{3} \ ) \biggr ] = 1.099\ 165\ 7...
\nonumber \\
 z_{q(5 \times \infty)} & = & \frac{2}{5}\Biggl [ 
\ln \Biggl ( \frac{9-\sqrt{5} + (70-18\sqrt{5})^{1/2} \ }{4} \Biggr ) + 
\ln \Biggl ( \frac{9+\sqrt{5} + (70+18\sqrt{5})^{1/2} \ }{4} \Biggr ) \Biggr ] 
\nonumber \\
 & = & 1.123\ 728\ 9...
\nonumber \\
 z_{sq(6 \times \infty)}&=& \frac{1}{6}\Biggl [
2 \ln \biggl ( \frac{3+\sqrt{5}\ }{2} \biggr ) + 
2 \ln \biggl ( \frac{5+\sqrt{21}\ }{2} \biggr ) + \ln(3+2\sqrt{2} \ ) \Biggr ] 
\nonumber \\
&=& 1.136\ 865\ 4...
\label{zsqn6} 
 \eeqs
 and
  \beqs
z_{tri(2 \times \infty)}&=& 2\ln2 = 1.386\ 294\ 3...
\nonumber  \\
 z_{tri(3 \times \infty)}&=& \frac{1}{3} \biggl [ -\ln 2 + 2\ln (7 + 3\sqrt{5} \ ) 
\biggr ] = 1.514\ 280\ 5...
\nonumber \\
 z_{tri(4 \times \infty)}&=&\frac{1}{2}\biggl [ 2\ln2 + \ln(3+\sqrt{7} \ ) \biggr ]
= 1.558\ 598\ 8...
\nonumber \\
  z_{tri(5 \times \infty)}&=& \frac{1}{5}\Biggl [ -7\ln2 + 
2\ln \Bigl ( 13-\sqrt{5}+(150-34\sqrt{5} \ )^{1/2} \Bigr ) \cr\cr
& & + 2\ln \Bigl ( 13+\sqrt{5}+(150+34\sqrt{5} \ )^{1/2} \Bigr ) \Biggr ] 
= 1.579\ 041\ 2...
\nonumber \\
 z_{tri(6 \times \infty)}&=&  \frac{1}{3}\biggl [ \ln(5+\sqrt{13} \ ) + 
\ln(7+3\sqrt{5} \ ) \biggr ] = 1.590\ 133\ 9... \ .
\label{ztrin6}
\eeqs
One observes that the values of $z$ 
are monotonically increasing in $N$, in accordance with
Theorem 2.1.  One also observes that 
$z_{{\cal L}(N_2 \times \infty)}$ converge reasonably quickly toward
the respective values $ z_{sq}$ in
(\ref{zsq}) and $z_{tri}$ in (\ref{ztri}).  
For example, $z_{sq(3 \times \infty)}$ and $z_{sq(6 \times \infty)}$ are, 
respectively, within 10 \% and 2.5 \%
of the value (\ref{zsq}) for the infinite square lattice.  Similarly,
$z_{tri(3 \times \infty)}$ and $z_{tri(6 \times \infty)}$ are, respectively,
within 6 \% and 1.5 \% of the value (\ref{ztri}).

\subsection{Connection with the Critical Ising Model and Dimers}

In this section we establish a result relating $z_{\cal L}$ for planar lattices
${\cal L}$ to the free energy of the Ising model on ${\cal L}$ at the critical
point.  We also remark on a connection of spanning trees with dimers for planar
lattices. We first state our result as a theorem.

\vspace{4mm}

\begin{flushleft}

Theorem 6.1.  

For  planar lattices ${\cal L}$ we have the identity
\beq
z_{\cal L} = a_{\cal L} + 2f_{\cal L}^c
\label{zfc}
\eeq
where $a_{\cal L}$ is a lattice-dependent constant, and $f_{\cal L}^c$ is the
(reduced) free energy of the Ising model on ${\cal L}$ at the critical point,
\vspace{5mm}

Proof.  We use the fact that the Ising model is the infinite 
bare quartic coupling ($\lambda\to\infty$) limit of the $\phi^4$ lattice field
theory \cite{wilson}. The partition function for the $\phi^4$ quantum field 
theory is 
\end{flushleft}
\beq
Z = \int_{-\infty}^\infty \Bigl [ \prod_i d\phi \Bigr ] e^{-S}
\label{pathintegral}
\eeq
where the action $S$ is an integral of the Lagrangian density with 
the quadratic part 
\beq
S_{quad} = \frac{1}{2} \int d^2x \biggl [ \sum_{j=1}^2 
\Bigl ( \frac{\partial \phi}{\partial x_j} \Bigr )^2 + m^2 \phi^2 \biggr ]\ . 
\label{quadlagrangian}
\eeq
After integrating by parts, the kinetic terms becomes 
$(1/2)\int d^2 x \phi [ -\partial^2 + m^2 ] \phi$, where $\partial^2$ 
is the Laplacian.  Further discretizing to a lattice ${\cal L}$, the integrand
in (\ref{quadlagrangian}) becomes the summand 
\beq
\frac{1}{2}\sum_{i,j} \phi_i Q_{ij} \phi_j + \frac{m^2}{2}
\sum_i \phi_i^2
\label{quadaction}
\eeq
where $Q_{ij}$ are elements of the Laplacian matrix 
 ${\bf Q}$ given by (\ref{matrixQ}).
 
At the critical point, because 
the phase transition is of second order with a divergent correlation length, 
the mass $m$ in (\ref{quadaction}) which is the inverse correlation length vanishes.
One is left 
simply with the term involving ${\bf Q}$.  Letting $\lambda \to \infty$, the 
functional integrals are now reduced to discrete sums over the Ising variables 
$\sigma_i = \pm 1$, and, from the Onsager solution and the correspondence with 
(\ref{mdef}), one finds  
\beq
f_{\cal L}^c = \frac{a_{\cal L}}{2} + \frac{1}{2\nu}
\int_{-\pi}^{\pi} {{d\theta_1}\over {2\pi}}
\int_{-\pi}^{\pi} {{d\theta_2}\over {2\pi}}
\ln  D(\theta_1,\theta_2) ,
\label{fc}
\eeq
where $a_{\cal L}$ is a lattice-dependent constant. 
This establishes  (\ref{zfc}) after using (\ref{main}).  $\Box$

\vspace{4mm}
We find the explicit results 
 \beqs
z_{sq}& =& -\ln 2 + 2 f_{sq}^c  \nonumber \\
z_{tri}& =& \ln (\sqrt{3}/2 ) + 2 f_{tri}^c  \nonumber \\
z_{hc}& =& -\ln(2\sqrt{3} \ ) + 2 f_{hc}^c \nonumber \\
z_{kag}&=&\frac{1}{2}\ln 3 -\frac{4}{3}\ln 2 - \frac{1}{3}\ln(2+\sqrt{3} \ ) 
+ 2 f_{kag}^c \ . 
\label{zfc2d}
\eeqs
The values of $f_{\cal L}^c$ in (\ref{zfc2d}) are well-known \cite{domb}. 

\vspace{4mm}

Finally, we remark that Temperley \cite{temp2} has established a bijection
between spanning trees on an $N\times N$ square lattice with free boundaries
and dimer configurations on a $(2N-1)\times (2N-1)$ square lattice with one
boundary site removed.  We shall not repeat the proof here, except to note that
this equivalence can be extended more generally to arbitrary planar graphs
\cite{wuu,propp}. Together with Theorem 6.1, this bijection implies a
connection between dimers and the critical Ising model, a connection which has
been observed by Fisher in a weaker form \cite{fisher1}.
 
\section{Discussion}

It is of interest to investigate how 
close the actual value of $z_{\cal L}$ is to the  Mckay-Fan-Yau 
upper bound (\ref{zmckay}).  To do this, we define 
for $\kappa$-regular lattices the ratio 
\beq
r_{{\cal L}} = \frac{z_{\cal L}}{\ln C_\kappa} \label{rl}
\eeq
where $C_\kappa$ is given by (\ref{ck}).  For  lattices 
which  are not $\kappa$-regular, we  compare $z_{\cal L}$ instead
  with the general upper bound (\ref{b1nonreg}) and consider the ratio 
\beq
r_{{\cal L}} = \frac{z_{\cal L}}{\ln \kappa_{eff}}
\eeq
Results are summarized in Table \ref{zarch}.
   
\begin{table}
\caption{\footnotesize{Values of $z_{\cal L}$ and $r_{\cal L}$ for 
different lattices ${\cal L}$.  }}
\begin{center}
\begin{tabular}{|c|c|c|c|c|c|c|}
\hline\hline
${\cal L}$ &d& $ \kappa_{\cal L}$ & $\kappa_{eff} $ & $\nu_{\cal L}$& 
$z_{\cal L}$ & $r_{\cal L}$ \\
\hline\hline
3-12-12 &2 & 3 & & 1/2 & 0.720\ 563\ 3...             & 0.861 \\
\hline
4-8-8 (Bathroom-tile) &2 & 3 & &1/2 & 0.786\ 684(1)            & 0.940 \\
\hline
Honeycomb  &2   & 3 & &1/2 & 0.807\ 664\ 8...            & 0.965 \\
\hline
Kagom\'e &2 & 4 &  & 1& 1.135\ 696\ 4...   & 0.933 \\
\hline
Diced &2 & & 4&1 & 1.135\ 696\ 4...  & 0.819 \\
\hline 
${\cal L}_2$ (Square) &2     & 4 & &1 & 1.166\ 243\ 6... & 0.959 \\
\hline
Union Jack &2&  & 6 & 2& 1.573\ 368(2)   & 0.878 \\
\hline
Triangular   &2 & 6 &  &2& 1.615\ 329\ 7...           & 0.955 \\
\hline
${\cal L}_3$ (Simple cubic)&3 & 6 &  &2 &   1.674\ 148\ 1(1) &  0.990 \\
\hline
${\cal L}_{bcc}$ (Body-centered cubic)&3 & 8  & &3 & 1.990\ 2(1) & 0.991 \\
\hline
${\cal L}_4$ &4 &  8 &  &3 &   2.000\ 0(5)  & 0.996 \\
\hline
${\cal L}_5$  &5& 10 &  &4 & 2.243          & 0.998 \\
\hline
${\cal L}_6$ &6 & 12 &  &5 & 2.437          & 0.999 \\
\hline
${\cal L}_{fcc}$ (Face-centered cubic)&3 & 12  & &5 & 2.354\ 4(4)  & 0.965 \\
\hline
${\cal L}_{bcc(4)}$ &4 & 16 & &7  &  2.732\ 3(1) & 0.998\\ 
\hline\hline
\end{tabular}
\end{center}
\label{zarch}
\end{table}

As is evident from Table 1, for regular lattices that we have studied, $z_{\cal
L}$ is a monotonically increasing function of the coordination number $
\kappa$.  We also observe that, for a fixed $\kappa$, the value of $z_{\cal L}$
increases with the spatial dimension of the lattice $z_{\cal L}$.  Examples are
(i) the triangular and simple cubic lattices (both with $\kappa =6)$, (ii) the
bcc and $d=4$ hypercubic lattices ($\kappa=8$), although the difference between
$z_{bcc}$ and $z_{{\cal L}_4}$ is very small, and (iii) the fcc and $d=6$
hypercubic lattices ($\kappa=12$).  Furthermore, our results indicates that in
two dimensions the square lattice is a little more densely connected than the
kagom\'e lattice, both of which have $\kappa=4$.  The ratio $r_{\cal L}$ is
observed in several cases to increase with $ \kappa$, but not in all cases; a
counterexample is $r=0.959$ for the square lattice ($\kappa=4$) and $r=0.955$
for the triangular lattice ($\kappa=6$).

\newpage
\noindent
{\bf Acknowledgments}

We are grateful to T. K. Lee for the hospitality at the National Center for
Theoretical Sciences, Taiwan, where this research was initiated. The research
of RS was supported in part at Stony Brook by the National Science Foundation
grant PHY-9722101 and at Brookhaven by the DOE contract
DE-AC02-98CH10886.\footnote{\footnotesize{Accordingly, the U.S. government
retains a non-exclusive royalty-free license to publish or reproduce the
published form of this contribution or to allow others to do so for
U.S. government purposes.}}  The research of FYW was supported in part by
National Science Foundation grants DMR-9614170 and DMR-9980440. The support of
the National Science Council, Taiwan, is also gratefully acknowledged.

\vfill
\eject
\end{document}